\documentclass[twocolumn]{aastex701}
\usepackage[utf8]{inputenc}

\usepackage{mathtools}
\usepackage{amssymb}
\usepackage{physics}
\usepackage{tensor}
\usepackage{cleveref}

\crefname{equation}{Equation}{Equations}

\newcommand{\Eq}[1]{\cref{eq:#1}}
\newcommand{\Eqns}[2]{\crefrange{eq:#1}{eq:#2}}

\newcommand{\App}[1]{Appendix \ref{app:#1}}

\newcommand{\sref}[1]{Section \ref{sec:#1}}

\newcommand{\Fig}[1]{Figure \ref{fig:#1}}

\newcommand{\code}[1]{{\ttfamily #1}}

\newcommand{\CC}[3]{\tensor{\Gamma}{^#1_{#2 #3}}}

\newcommand{\cd}[1]{\tensor{\grad}{#1}}
\newcommand{\pd}[1]{\tensor{\partial}{#1}}

\newcommand{\gcc}{\text{g cm}^{-3}}

\newcommand{\affiliationNCCS}{\affiliation{National Center for Computational Sciences, Oak Ridge National Laboratory, Oak Ridge, TN 37831}}

\newcommand{\affiliationPHYS}{\affiliation{Physics Division, Oak Ridge National Laboratory, Oak Ridge, TN 37831}}
\newcommand{\affiliationUTK}{\affiliation{Department of Physics and Astronomy, University of Tennessee, Knoxville, TN 37996}}

\begin{document}


\title{Approximating General Relativity in Core-Collapse Supernova Simulations\footnote{This manuscript has been authored in part by UT-Battelle, LLC, under contract DE-AC05-00OR22725 with the US Department of Energy (DOE). The US government retains and the publisher, by accepting the article for publication, acknowledges that the US government retains a nonexclusive, paid-up, irrevocable, worldwide license to publish or reproduce the published form of this manuscript, or allow others to do so, for US government purposes. DOE will provide public access to these results of federally sponsored research in accordance with the DOE Public Access Plan (http://energy.gov/downloads/doe-public-access-plan).}}

\shortauthors{Fromm et al.}
\shorttitle{Approx. GR in CCSN Simulations}

\author[0000-0002-3591-123X]{Steven A.~Fromm}
\affiliationNCCS
\email{frommsa@ornl.gov}

\author[0000-0001-5869-8542]{Vassilios Mewes}
\affiliationNCCS
\email{mewesv@ornl.gov}

\author[0000-0002-5358-5415]{O.~E.~Bronson Messer}
\affiliationNCCS
\affiliationPHYS
\affiliationUTK
\email{bronson@ornl.gov}

\author[0000-0002-5231-0532]{Eric J.~Lentz}
\affiliationUTK
\affiliationPHYS
\email{elentz@utk.edu}

\author[0000-0002-9481-9126]{W.~Raphael Hix}
\affiliationPHYS
\affiliationUTK
\email{raph@utk.edu}

\author[0000-0003-3023-7140]{J.~Austin Harris}
\affiliationNCCS
\email{harrisja@ornl.gov}

\correspondingauthor{Steven A.~Fromm}
\email{frommsa@ornl.gov}

\begin{abstract}
    We present formulations of effective potentials suitable for approximating general relativistic effects in Newtonian simulations of core-collapse supernovae.  Assuming a spherically symmetric spacetime and a stress-energy tensor that includes both fluid and neutrino contributions, Eulerian and Lagrangian projections of the Einstein equations are made to determine general relativistic corrections to the Newtonian gravitational potential.  We implement the effective potentials in both the {\ttfamily Chimera} and {\ttfamily Flash-X} codes, and perform a series of adiabatic and core collapse simulations.  The results are compared to Newtonian and fully general relativistic simulations, as well as another widely used effective potential formulation.  We find close agreement between our new effective potentials and the fully general relativistic results from multiple other codes.
\end{abstract}



\section{Introduction}\label{sec:introduction}

A core-collapse supernova (CCSN) results from the death of a massive star (${\gtrsim}10 M_\odot$) when its degenerate iron core exceeds the Chandrasekhar mass.
The collapse of the core that follows results in the formation of a shock wave, which may disrupt the star, and a neutron star, at least for a brief time before the neutron star itself exceeds its maximum mass and collapses into a black hole.
The formation of the neutron star, in turn, results in the release of ${\sim} 10^{53}$ erg of neutrinos, carrying off the gravitational binding energy of the newborn neutron star, and any matter from the progenitor star that subsequently accretes onto the neutron star, over the course of several seconds.
The progress of the shock wave through the star is impeded by further neutrino emission and nuclear dissociation, which enervates the shock and causes it to stall.
A successful CCSN requires re-energizing the shock wave, which is generally accomplished by capturing a small fraction of the escaping neutrinos in the matter behind the shock. 
Thus, CCSNe are ultimately gravitationally powered, with neutrinos serving as the intermediary.
For a review of the relevant physics see, for instance, \cite{Mezzacappa2005,Janka2007,Burrows2013,Foglizzo2015,Couch2017,Mueller2020,Mezzacappa2020,Boccioli2024,Janka2025}.

The effects of general relativity (GR) become increasingly important at the high densities and compactness present in the cores of these collapsing stars \citep{Bruenn2001}, where Newtonian gravitation can no longer fully describe the dynamical spacetime.
These conditions not only necessitate a GR description of the spacetime, but also of the evolving matter and neutrinos.
Including GR effects in CCSN simulations can produce a marked difference compared to Newtonian simulations.
Improved neutrino heating rates and increased explodability were observed by \citet{Oconnor2018a} and \citet{Bruenn2001} for simulations including both full and approximate GR.  
In some instances, only a full GR treatment of the spacetime, fluid, and radiation lead to an explosion, whereas Newtonian and approximate GR simulations fail to explode, as seen by \citet{Muller2012}.
The more compact neutron star that results from GR both boosts the gravitational binding energy available to radiate and reduces the area of the surface of the neutron star, which serves as the neutrino radiating surface or neutrinosphere.  
This, in turn, hardens the neutrino spectrum, enhancing the neutrino heating rate behind the shock.

However, full GR descriptions of the spacetime, hydrodynamics, and neutrino radiation transport in multidimensional CCSN simulations (see, e.g.~\cite{Mueller2010,Roberts2016,Kuroda2016,Rahman2019,Kuroda2020,Akaho2021}) are computationally challenging, bordering on being impractical alongside the high-fidelity physics often included in these simulations, e.g., spectral radiation transport and detailed fluid composition evolution via nuclear reaction networks.
Instead, approximating GR effects to augment non-relativistic or special relativistic hydrodynamics and radiation transport solvers offers a more tractable solution for CCSN simulations.
One method for approximating these GR effects, pioneered by \citet{Rampp2002} and later refined by \citet{Marek2006}, exploits the relative sphericity of the dense compact core to replace the monopole term of the Newtonian gravitational potential with a spherically symmetric potential based on the Tolman-Oppenheimer-Volkoff (TOV) equations \citep{Tolman1939,Oppenheimer1939} that accounts for both rest mass and energy densities in the gravitational mass.
This method, in particular the `Case A' modification presented in \citet{Marek2006}, has seen widespread and continued use in CCSN simulations, e.g., \citet{Buras2006,Muller2012,Kotake2018,Oconnor2018a,Skinner2019,Bruenn2020}, producing favorable comparisons with full GR results. 

This paper will present a new method for approximating GR effects in both Lagrangian and Eulerian Newtonian hydrodynamics codes via effective potentials that follows a similar approach to the one taken in \citet{Rampp2002} and \citet{Marek2006}.
In \sref{formulation} we will present an overview of its formulation, and in \sref{results} we will demonstrate its ability to capture the effects of GR in adiabatic collapse, CCSN, and isolated neutron star scenarios.
Geometric units with $G = c = M_\odot = 1$ will be used throughout this paper unless otherwise stated explicitly.
The lower-case Latin indices $a,b,c,\ldots$ will represent full spacetime indices, while $i,j,k,\ldots$ will represent spatial indices.
The Einstein summation convention, i.e., $\tensor{u}{_a}\tensor{u}{^a} \equiv \sum_{a=0}^{3}\tensor{u}{_a}\tensor{u}{^a}$, will be assumed for all tensorial equations.
Pairs of indices surrounded by parentheses indicate symmetrization, e.g., $\tensor{T}{^{(ab)}} \coloneq \frac{1}{2}\qty(\tensor{T}{^{ab}} + \tensor{T}{^{ba}})$.

\section{Formulation}\label{sec:formulation}

The spacetime of a spherically symmetric self-gravitating star in equilibrium is described by the line element \citep{Gourgoulhon1991}
\begin{align}\label{eq:tov_ds2}
    \dd{s}^2 = -\alpha^2 \dd{t}^2 + X^2\dd{r}^2 + r^2\qty(\dd{\theta}^2 + \sin^2\theta\dd{\varphi}^2),
\end{align}
where $\alpha \equiv \alpha(r)$ and $X \equiv X(r)$ depend only on the radial coordinate $r$, and relate to the metric potential $\Phi(r)$ and enclosed gravitational mass $m(r)$ as
\begin{align}
    \alpha(r) &= e^{\Phi(r)}\label{eq:tov_alpha},\\
    X(r) &= \qty(1 - \frac{2m(r)}{r})^{-1/2}\label{eq:tov_X}.
\end{align}
The structures of the star and spacetime are found by solving the Einstein equations \citep{misner2017}
\begin{align}\label{eq:einstein_eq}
    \tensor{R}{_{ab}} - \frac{1}{2}R\tensor{g}{_{ab}} = 8\pi\tensor{T}{_{ab}},
\end{align}
where $\tensor{g}{_{ab}} = \text{diag}\qty(-\alpha^2,X^2,r^2,r^2\sin^2\theta)$ is the spacetime metric associated with the line element in \Eq{tov_ds2}, $\tensor{R}{_{ab}}$ is the Ricci tensor, $R = \tensor{g}{^{ab}}\tensor{R}{_{ab}}$ is the Ricci scalar, and $\tensor{T}{_{ab}}$ is the stress-energy tensor of the star.
For CCSN applications we will use a stress-energy tensor of the form
\begin{align}\label{eq:Tab}
    \tensor{T}{^{ab}} = T^{ab}_\text{fluid} + \mathcal{T}^{ab}_\nu,
\end{align}
where $T^{ab}_\text{fluid}$ and $\mathcal{T}^{ab}_\nu$ represent the contributions of the fluid and neutrinos, respectively.  We assume an ideal fluid with a stress-energy tensor
\begin{align}\label{eq:Tab_fluid}
    T^{ab}_\text{fluid} = \rho_0 h \tensor{u}{^a} \tensor{u}{^b} + p \tensor{g}{^{ab}},
\end{align}
where $\rho_0$ is the rest-mass density, $p$ is the fluid pressure, $h = \qty(\rho + p)/\rho_0$ is the specific enthalpy, $\rho = \rho_0\qty(1+\epsilon)$ is the fluid energy density, and $\epsilon$ is the specific internal energy, as measured by a co-moving observer with four-velocity $\tensor{u}{^a}$.
The neutrino stress-energy tensor takes the form
\begin{align}\label{eq:Tab_nu}
    \mathcal{T}^{ab}_\nu = \mathcal{J}\tensor{u}{^a}\tensor{u}{^b} + 2\tensor{\mathcal{H}}{^{(a}}\tensor{u}{^{b)}} + \tensor{\mathcal{K}}{^{ab}},
\end{align}
where the energy density $\mathcal{J}$, momentum density $\tensor{\mathcal{H}}{^a}$, and pressure tensor $\tensor{\mathcal{K}}{^{ab}}$ are measured in the frame of the co-moving observer.
These quantities represent the total contributions of all neutrino species and energies, i.e.,
\begin{align}\label{eq:gray_moments}
    \qty{\mathcal{J},\tensor{\mathcal{H}}{^a},\tensor{\mathcal{K}}{^{ab}}} = \sum_s \int_0^\infty\dd{\varepsilon}\qty{\mathcal{J},\tensor{\mathcal{H}}{^a},\tensor{\mathcal{K}}{^{ab}}}_\qty(s,\varepsilon),
\end{align}
where $s \in \qty{\nu_e, \bar{\nu}_e, \ldots}$ labels the neutrino species, and $\varepsilon$ is the neutrino energy in the frame of the co-moving observer.

Projecting \Eq{einstein_eq} along and orthogonal to the four-velocity $u^a$ we arrive at (see \App{derivations} for details)
\begin{align}
    \dv{\ln X}{r} &= 4\pi r X^2 \qty(\rho + \mathcal{J}) - \frac{1}{2r}\qty(X^2 - 1)\label{eq:dX_dr},\\
    \dv{\ln \alpha}{r} &= 4\pi r X^2 \qty(p + \mathcal{K}) + \frac{1}{2r}\qty(X^2 - 1)\label{eq:dalpha_dr},
\end{align}
where we use $\mathcal{K} \equiv X^{-2}\tensor{\mathcal{K}}{_{rr}}$.
Using \Eqns{tov_alpha}{tov_X} in \Eqns{dX_dr}{dalpha_dr} yields equations for the gravitational mass and metric potential
\begin{align}
    \dv{m}{r} &= 4\pi r^2\qty(\rho + \mathcal{J})\label{eq:dm_dr},\\
    \dv{\Phi}{r} &= X^2\qty[\frac{\vphantom{2}m}{r^2} + 4\pi r \qty(p + \mathcal{K})]\label{eq:dPhi_dr}.
\end{align}
As anticipated, \Eq{dm_dr} and \Eq{dPhi_dr} extend the TOV equations to include a neutrino contribution.

We then identify the metric potential $\Phi(r)$ as a spherically symmetric gravitational potential associated with the spacetime of the star.
Using the stress-energy tensor in \Eq{Tab} to solve $\cd{_a}\tensor{T}{^a_r} = 0$ we find that
\begin{align}\label{eq:hydrostatic_eq}
    \dv{p}{r} + \qty(\rho + p)\dv{\Phi}{r} = -\qty[\dv{\mathcal{K}}{r} + \qty(\mathcal{J} + \mathcal{K})\dv{\Phi}{r}].
\end{align}
In the absence of neutrinos, \Eq{hydrostatic_eq} reduces to the standard TOV hydrostatic equilibrium condition for a relativistic star, demonstrating that $\Phi(r)$ plays the role of a gravitational potential.
To determine an overall effective potential $\bar{\Phi}_\text{L}(r)$, we take a similar approach as \citet{Marek2006} and replace the spherically symmetric part of the Newtonian gravitational potential with the metric potential
\begin{align}\label{eq:Phi_eff_full}
    \bar{\Phi}_\text{L}(\vb{r}) = \Phi_N(\vb{r}) - \Phi_0(r) + \Phi(r),
\end{align}
where $\vb{r}$ is the coordinate position vector with radius $r$, and $\Phi_N(\vb{r})$ is the full Newtonian potential with its spherically symmetric part given by
\begin{align}\label{eq:Phi_0}
    \Phi_0(r) = \int_\infty^r \dd{r}' \frac{m_0}{r'^2},
\end{align}
with an enclosed rest-mass
\begin{align}\label{eq:m_0}
    m_0(r) = 4\pi\int_0^r \dd{r}' r'^2\rho_0.
\end{align}

Although \Eq{dPhi_dr} can be used to directly calculate $\Phi(r)$ in \Eq{Phi_eff_full}, we find it useful to modify this equation to use a rescaled gravitational mass
\begin{align}\label{eq:mbar}
    \bar{m} = X^{-2}m,
\end{align}
such that \Eq{dPhi_dr} becomes
\begin{align}\label{eq:dPhi_dr_mbar}
    \dv{\Phi}{r} &= X^2\qty[\frac{\vphantom{2}\bar{m}}{r^2} + 4\pi r \qty(p + \mathcal{K})]\nonumber\\
    &= \frac{m}{r^2} + 4\pi r X^2 \qty(p + \mathcal{K}).
\end{align}
Integrating \Eq{dm_dr} shows that the gravitational mass takes the form
\begin{align}\label{eq:m_0_plus_dm}
    m(r) &= 4\pi\int_0^r \dd{r'}r'^2 \qty(\rho + \mathcal{J})\nonumber\\
    &= 4\pi\int_0^r \dd{r'}r'^2 \rho_0 + 4\pi\int_0^r \dd{r'}r'^2 \qty(\rho_0\epsilon + \mathcal{J})\nonumber\\
    &= m_0(r) + \delta m(r),
\end{align}
where $\delta m(r)$ contains GR corrections to the rest-mass to obtain the gravitational mass.
Similarly, integrating \Eq{dPhi_dr_mbar} shows that
\begin{align}\label{eq:Phi_deltaPhi}
    \Phi(r) &= \int_\infty^r \dd{r}'\qty[\frac{m}{r'^2} + 4\pi r'X^2 \qty(p + \mathcal{K})]\nonumber\\
    &= \int_\infty^r \dd{r}'\frac{m_0}{r'^2} + \int_\infty^r \dd{r}'\qty[\frac{\delta m}{r'^2} + 4\pi r'X^2 \qty(p + \mathcal{K})]\nonumber\\
    &= \Phi_0(r) + \delta \Phi(r),
\end{align}
where the $\delta \Phi(r)$ term once again represents GR corrections to the spherically symmetric Newtonian potential.
Using \Eq{Phi_deltaPhi} in \Eq{Phi_eff_full} then gives the effective potential as
\begin{align}\label{eq:Phi_effective}
    \bar{\Phi}_\text{L}(\vb{r}) = \Phi_N(\vb{r}) + \delta\Phi(r).
\end{align}
For the effective potential in \Eq{Phi_effective} we only need to compute the spherically symmetric GR corrective terms to add to an existing Newtonian potential calculation.

We make additional projections of \Eq{einstein_eq} in the frame of an Eulerian observer with a four-velocity $\tensor{n}{^a}$ such that the Lorentz factor between this observer and the co-moving observer is $W = -\tensor{n}{_a}\tensor{u}{^a}$.
These projections yield (see \App{derivations} for details)
\begin{align}
    \dv{\ln X}{r} &= 4\pi r X^2 \qty(\rho_0hW^2 - p + \mathcal{E}) - \frac{1}{2r}\qty(X^2 - 1)\label{eq:dX_dr_E},\\
    \dv{\ln \alpha}{r} &= 4\pi r X^2 \qty(\rho_0hW^2v^2 + p + \mathcal{P}) + \frac{1}{2r}\qty(X^2 - 1)\label{eq:dalpha_dr_E},
\end{align}
where $v$ is the fluid radial velocity, and $\mathcal{E}$ and $\mathcal{P}$ are the neutrino energy density and pressure, respectively, as measured by the Eulerian observer.
Substituting the definitions in \Eqns{tov_alpha}{tov_X} into \Eqns{dX_dr_E}{dalpha_dr_E} then gives
\begin{gather}
    \Phi_\text{E}(r) = \int_\infty^r\dd{r'}X_\text{E}^2\qty[\frac{m_\text{E}}{r'^2} + 4\pi r'\qty(\rho_0hW^2v^2 + p + \mathcal{P})]\label{eq:Phi_E},\\
    m_\text{E}(r) = 4\pi\int_0^r\dd{r'}r'^2\qty(\rho_0hW^2 - p + \mathcal{E})\label{eq:m_E},
\end{gather}
where $X_\text{E}$ is the metric function in \Eq{tov_X} evaluated with $m_\text{E}$.
\Eqns{Phi_E}{m_E} take a similar form as the potential and mass in the radial-gauge polar-slicing formulations used in \citet{Gourgoulhon1991,OConnor2010,OConnor2015}.
The effective potential in the Eulerian frame then takes the form
\begin{align}\label{eq:Phibar_E}
    \bar{\Phi}_\text{E}(\vb{r}) = \Phi_N(\vb{r}) - \Phi_0(r) + \Phi_\text{E}(r).
\end{align}

\section{Results}\label{sec:results}

We have implemented the effective gravitational potentials described in \sref{formulation} in both the \code{Chimera} \citep{Bruenn2020} and \code{Flash-X} \citep{Dubey2022} codes.
To assess the quality of the effective potentials, we perform a series of collapse and isolated neutron star evolutions, and we compare our results to the fully GR results from the \code{AGILE-BOLTZTRAN} \citep{Liebendorfer2004}, \code{GR1D} \citep{OConnor2010,OConnor2015}, and \code{SphericalNR} \citep{Mewes2018,Mewes2020} codes.
We also compare our effective potential with the `Case A' GR effective potential from \citet{Marek2006}, hereafter referred to as \code{GREP}; see \App{grep} for an overview of this potential and how it differs from the effective potentials presented in \sref{formulation}.

\subsection{Codes used in this comparison}\label{sec:codes}
\subsubsection{\code{Chimera}}\label{sec:chimera}

\code{Chimera} utilizes a Lagrangian-plus-remap hydrodynamics solver and multigroup flux-limited diffusion (MGFLD) neutrino radiation transport solver.
We have implemented the effective potential $\bar{\Phi}_\text{L}$ in \Eq{Phi_effective} as an alternative to the existing \code{GREP} implementation.
Both potentials are calculated on a spherically-averaged grid of primitive fluid and neutrino variables.
\code{Chimera} also computes the lapse function $\alpha$ in terms of the effective potentials for use in the MGFLD solver.
We use \code{Chimera} for both the adiabatic collapse and CCSN simulations presented later in this section.
For CCSN simulations, \code{Chimera} evolves four neutrino species, $\nu_e$, $\bar{\nu}_e$, and combined heavy neutrino flavors $\nu_x = \qty{\nu_\mu,\nu_\tau}$ and $\bar{\nu}_x = \qty{\bar{\nu}_\mu,\bar{\nu}_\tau}$, using 20 logarithmically spaced energy bins centered at $4$ to $250$~MeV.

\subsubsection{\code{Flash-X}}\label{sec:flashx}

\code{Flash-X} utilizes an Eulerian finite-volume hydrodynamics solver with adaptive mesh refinement.
For this work, we have implemented the effective potential $\bar{\Phi}_\text{E}$ in \Eq{Phibar_E}, and have ported over the implementation of \code{GREP} in \code{FLASH} \citep{Oconnor2018a} to \code{Flash-X}.
Similarly to \code{Chimera}, both effective potentials are computed on a spherically-averaged grid of primitive fluid variables.
\code{Flash-X} includes neutrino radiation transport via coupling to \code{thornado} \citep{Endeve2026} which implements a discontinuous Galerkin discretization of a $\order{v/c}$ two-moment number conservative scheme.
We are currently working on including the effective potential contributions via an effective lapse in \code{thornado}, so here we only include adiabatic collapse and isolated neutron star simulations for \code{Flash-X}.

\subsubsection{\code{GR1D}}\label{sec:gr1d}

\code{GR1D} includes Eulerian finite-volume GR hydrodynamics and two-moment multigroup neutrino radiation transport solvers.  All fluid and radiation variables are evolved on a fixed spherically symmetric grid.
Metric functions are not dynamically evolved, but computed directly from the fluid and radiation variables.
For CCSN simulations, \code{GR1D} evolves three neutrino species, $\nu_e$, $\bar{\nu}_e$, and a combined heavy neutrino $\nu_x = \qty{\nu_\mu, \bar{\nu}_\mu, \nu_\tau, \bar{\nu}_\tau}$, using 18 logarithmically spaced energy bins centered at $1$ to ${\sim}280.5$~MeV.
\code{GR1D} can be run with Newtonian hydrodynamics and radiation transport, with and without an effective potential, but we only use its fully GR solvers for both the adiabatic collapse and CCSN simulations presented later in this section.

\subsubsection{\code{AGILE-BOLTZTRAN}}\label{sec:agile_boltztran}

\code{AGILE-BOLTZTRAN} is a combination of the GR hydrodynamics code \code{AGILE} \citep{Liebendorfer2002} and the neutrino transport code \code{BOLTZTRAN} \citep{Mezzacappa1993,Mezzacappa1999,Liebendorfer2004}.
\code{AGILE} solves the complete GR spacetime and hydrodynamics equations implicitly in spherical symmetry on a dynamic, moving grid.
\code{BOLTZTRAN} solves the GR neutrino Boltzmann equation using the method of discrete ordinates ($S_N$) with a  Gauss-Lobatto quadrature.
For the CCSN results presented below, we use 6-point angular quadrature to resolve the angular distribution of the radiation field and 20 logarithmically-spaced energy groups centered from 3 to 300~MeV.
We do not include any physics to distinguish between muon- and tau-flavored leptons, we use the combined species $\nu_x = \qty{\nu_\mu,\nu_\tau}$ and $\bar{\nu}_x = \qty{\bar{\nu}_\mu,\bar{\nu}_\tau}$.

\subsubsection{\code{SphericalNR}}\label{sec:sphericalnr}

\code{SphericalNR} is a framework within the \code{Einstein Toolkit}~\citep{ETK2012,EinsteinToolkit:2025_05} 
to solve the Einstein field equations coupled to the equations of GR magnetohydrodynamics in spherical coordinates without symmetry assumptions. For the simulation in Section~\ref{sec:tov}, \code{SphericalNR} was run with a ``fisheye'' radial coordinate~\citep{Ji2023} and
$\vartheta=4,\varphi=2$ cells in the angular coordinates. A purely hydrodynamic evolution was ensured by running
with a vanishing magnetic field.

\subsection{Adiabatic Collapse}\label{sec:adiabatic}

We first assess the quality of the effective potentials in a purely hydrodynamical adiabatic collapse by running simulations in \code{Chimera} and \code{Flash-X} with a fully Newtonian potential, \code{GREP}, and the effective potentials in \sref{formulation}.
We evolve the $15M_\odot$ progenitor from \citet{Woosley2007} with each code, using the SFHo equation of state (EOS) \citep{Steiner2013}.
For comparison, we also run full GR simulations using the same progenitor and EOS in both \code{GR1D} and \code{AGILE-BOLTZTRAN}.
All simulations are run using one-dimensional spherically symmetric grids.

\begin{figure}[ht!]
   \centering
   \includegraphics[width=0.9\columnwidth]{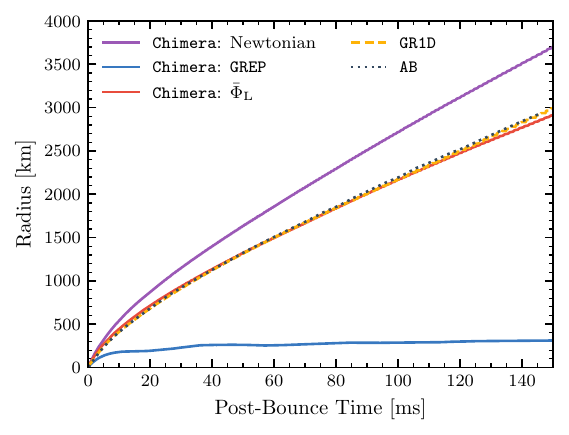}
   \caption{\label{fig:shock_adiabatic_chimera}Shock expansion results from the \code{Chimera} adiabatic collapse simulations of the $15M_\odot$ progenitor from \citet{Woosley2007}.  The new effective potential $\bar{\Phi}_\text{L}$ (solid red line) shows close agreement with full GR simulations in \code{GR1D} (dashed yellow line) and \code{AGILE-BOLTZTRAN} (AB, dotted black line), all of which produce a slower shock expansion than the Newtonian potential (solid purple line).  However, \code{GREP} (solid blue line) shows a qualitatively different outcome of a shock that stalls ${\sim}10$ ms post-bounce.}
\end{figure}

\begin{figure}[ht!]
   \centering
   \includegraphics[width=0.9\columnwidth]{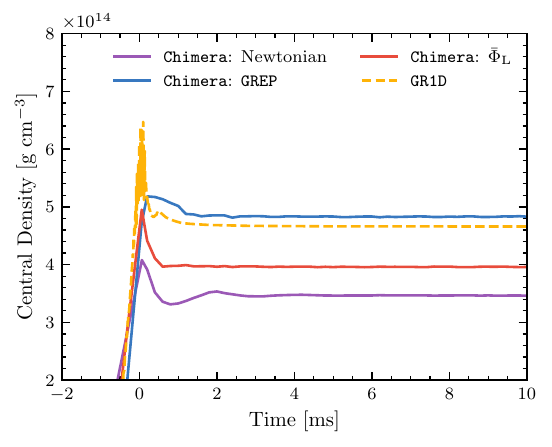}
   \caption{\label{fig:rho_c_adiabatic_chimera}Central density evolution near bounce from the \code{Chimera} adiabatic collapse simulations of the $15M_\odot$ progenitor from \citet{Woosley2007}.  \code{GREP} (solid blue line) reaches a post-bounce central density comparable to the full GR simulation in \code{GR1D} (dashed yellow line).  The Newtonian potential (solid purple line) produces a marginally lower post-bounce central density, and the new effective potential $\bar{\Phi}_\text{L}$ (solid red line) falls in between.}
\end{figure}

\begin{figure}[ht!]
   \centering
   \includegraphics[width=0.9\columnwidth]{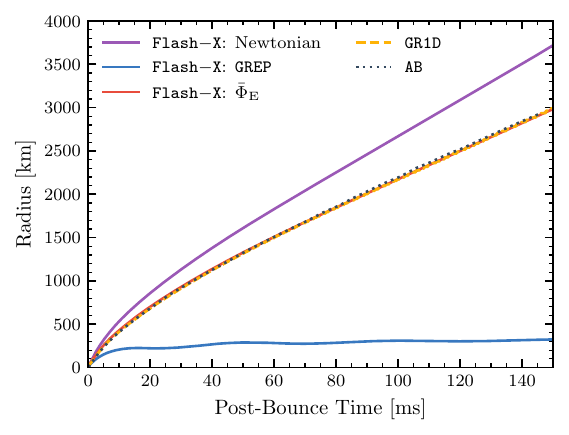}
   \caption{\label{fig:shock_adiabatic_flashx}Same as \Fig{shock_adiabatic_chimera}, but for \code{Flash-X} Newtonian and effective potential simulations, and with the Eulerian effective potential ($\bar{\Phi}_\text{E}$, red solid line).}
\end{figure}

\begin{figure}[ht!]
   \centering
   \includegraphics[width=0.9\columnwidth]{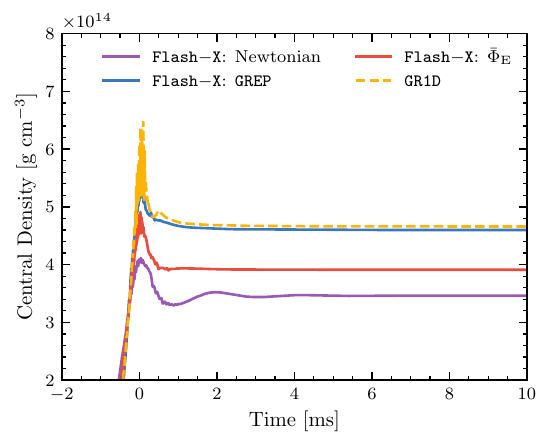}
   \caption{\label{fig:rho_c_adiabatic_flashx}Same as \Fig{rho_c_adiabatic_chimera}, but for \code{Flash-X} Newtonian and effective potential simulations, and with the Eulerian effective potential ($\bar{\Phi}_\text{E}$, red solid line).}
\end{figure}

The results for post-bounce shock radii of the adiabatic collapse simulations are shown in \Fig{shock_adiabatic_chimera} and \Fig{shock_adiabatic_flashx}, while the central densities near bounce are shown in \Fig{rho_c_adiabatic_chimera} and \Fig{rho_c_adiabatic_flashx}.
We observe that the purely Newtonian potential produces the fastest shock expansion, the effective potentials from \sref{formulation} expands more slowly, but remain close to the full GR results, and the simulations using \code{GREP} show a qualitatively different outcome of a shock that quickly stalls approximately $10$ ms post-bounce.
We observe similar post-bounce central densities for each set of simulations between the codes: the Newtonian simulations settle in at the lowest values, the \code{GREP} values are approximately in agreement with the full GR result in \code{GR1D}, and the $\bar{\Phi}_\text{L}$ and $\bar{\Phi}_\text{E}$ potentials produce central densities in between.  
The difference in behavior between \code{GREP} and our new effective potential likely reflects different priorities in the choice of GR behavior to match.

\subsection{CCSN}\label{sec:ccsn}

We next assess the quality of the effective potentials in CCSN simulations by running simulations in \code{Chimera} using a fully Newtonian potential, \code{GREP}, and the effective potential $\bar{\Phi}_\text{L}$ described in \sref{formulation}.
Again, we evolve the $15M_\odot$ progenitor from \citet{Woosley2007} with each code while using the SFHo EOS \citep{Steiner2013}.
For the neutrino-matter interactions, we include the microphysics described in \citet{bruenn1985} as well as
nucleon-nucleon Bremsstrahlung~\citep{hannestad1998}.
For comparison, we also run full GR simulations in \code{GR1D} and \code{AGILE-BOLTZTRAN} using the same progenitor and EOS.
We use a custom-built \code{NuLib} \citep{OConnor2015} table to match the microphysics as closely as possible in the \code{GR1D} simulation.
All simulations are run using one-dimensional spherically-symmetric grids.

\begin{figure}[ht!]
   \centering
   \includegraphics[width=0.9\columnwidth]{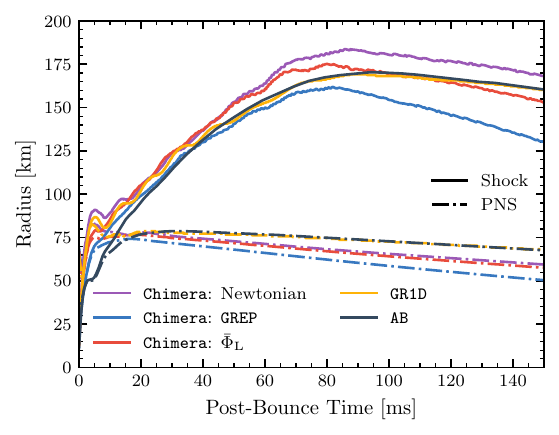}
   \caption{\label{fig:shock_ccsn_chimera}Shock and PNS radius results from \code{Chimera} CCSN simulations of the $15M_\odot$ progenitor from \citet{Woosley2007}.  With the new effective potential $\bar{\Phi}_\text{L}$ (solid red line) the shock expands more slowly than with the Newtonian potential (solid purple line), but faster than \code{GREP} (solid blue line).  A similar ordering is observed in the PNS radius (dashed-dotted lines).  The results from full GR simulations in \code{GR1D} (yellow lines) and \code{AGILE-BOLTZTRAN} (AB, black lines) are included for reference.}
\end{figure}

\begin{figure}[ht!]
   \centering
   \includegraphics[width=0.9\columnwidth]{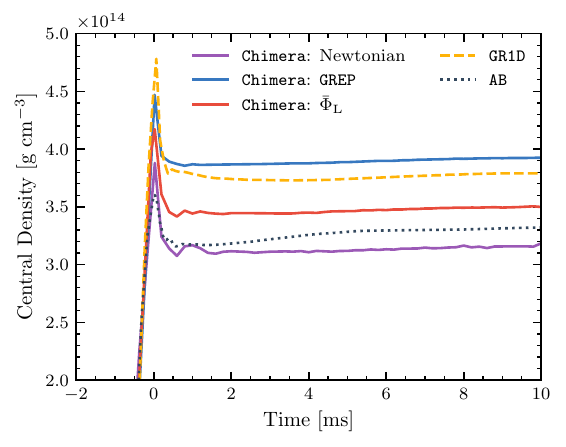}
   \caption{\label{fig:rho_c_ccsn_chimera}Central density evolution near bounce from \code{Chimera} CCSN simulations of the $15M_\odot$ progenitor from \citet{Woosley2007}.  \code{GREP} (blue line) reaches the highest post-bounce central density, slightly above the \code{GR1D} result (yellow dashed line).  The new effective potential $\bar{\Phi}_\text{L}$ (red line) and the Newtonian potential (purple line) are consecutively lower, with the \code{AGILE-BOLTZTRAN} (AB, black dotted line) falling in between.}
\end{figure}

The results of each simulation for the post-bounce shock and proto-neutron star (PNS) radii are shown in \Fig{shock_ccsn_chimera} and for the central density evolution near bounce in \Fig{rho_c_ccsn_chimera}.
Again, we observe similar results as in the adiabatic collapse simulations with the Newtonian potential producing the fastest and furthest initial shock expansion, \code{GREP} producing the slowest initial shock expansion, and the results for the new effective potential described in \sref{formulation} falling in between.
The PNS radii also follow a similar ordering, but we note that in all cases these show an increasing deviation from the full GR results.
Likewise, the post-bounce central densities follow a similar pattern as the adiabatic collapse results, with the $\bar{\Phi}_\text{L}$ effective potential reaching a value in between the Newtonian and GR results, while the \code{GREP} result reaches a similar central density as the full GR results.
We note that while \code{Chimera} produces a similar shock expansion as in the full GR results, the differences in radiation transport methods between \code{Chimera} and both \code{GR1D} and \code{AGILE-BOLTZTRAN}, as well as minor differences in the included microphysics, prevent any definitive comparison of the effective potentials with the full GR results.

\subsection{Isolated Neutron Star}\label{sec:tov}

As a third test of the effective potentials we simulate an isolated neutron star's migration from an unstable branch central density (see, e.g.,~\cite{Font2002,Cordero2009}).
To set up this test, we solve the TOV equations (\Eqns{dm_dr}{hydrostatic_eq} without the neutrino-specific terms) using a central density of $\rho_c = 4.93\times10^{15}~\gcc$ and a polytropic equation of state $p\qty(\rho_0) = K \rho_0^\gamma$ with $\gamma = 2$ and $K = 1.455\times10^5$ (cgs units), resulting in a $1.44~M_\odot$ neutron star.
Following \citet{Marek2006}, we generate an initial solution for the same central density using the modified `Case A' effective potential in place of the standard TOV potential for its respective simulation.
We perform simulations of these initial models for each of their respective effective potentials in \code{Flash-X} to compare against the full GR results from \code{SphericalNR}. The initial data for \code{SphericalNR} is generated
with the \code{Hydro\_RNSID} thorn based on the \code{RNS} code~\citep{Stergioulas1995}.

\begin{figure}[ht!]
   \centering
   \includegraphics[width=0.9\columnwidth]{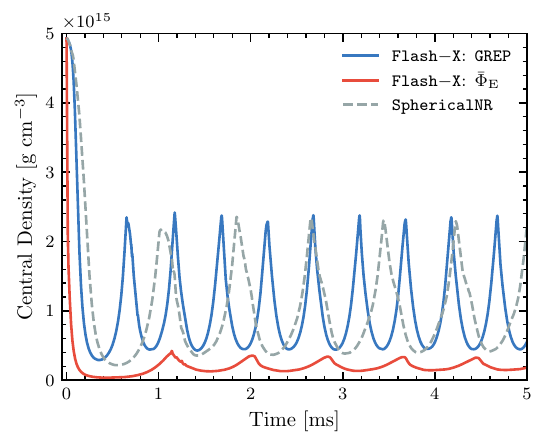}
   \caption{\label{fig:tov_rho_c_flashx}\code{Flash-X} central density evolution for the isolated neutron star simulations.  Both \code{GREP} (blue solid line) and the $\bar{\Phi}_\text{E}$ (red solid line) effective potential migrate to a new stably evolving state, with the $\bar{\Phi}_\text{E}$ potential migrating to a lower central density than the GR result from \code{SphericalNR} (gray dashed line).}
\end{figure}

\begin{figure}[ht!]
   \centering
   \includegraphics[width=0.9\columnwidth]{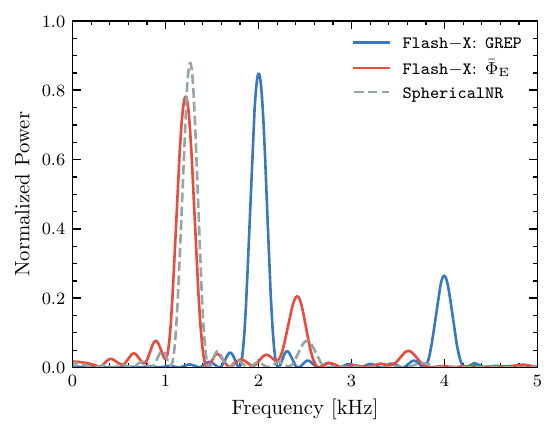}
   \caption{\label{fig:tov_freqs_flashx}\code{Flash-X} frequency spectrum of the central density oscillations for the isolated neutron star simulations.  The new effective potential $\bar{\Phi}_\text{E}$ (red solid line) and the GR results from \code{SphericalNR} (gray dashed line) oscillate at about ${\sim}1.2$ kHz around their migrated central densities, while the \code{GREP} (blue solid line) result reaches a central density that oscillates at ${\sim}2$ kHz.}
\end{figure}

For all simulations we see a successful migration of the initial unstable branch solution to a new equilibrium solution at a lower central density.
The results for the central density shown in \Fig{tov_rho_c_flashx} show that the \code{GREP} simulation evolves to a similar central density as the full GR result, while similar to the collapse simulations we observe the effective potential $\Phi_\text{E}$ evolves to a lower central density.
The frequency spectrum of the central density oscillations after the migration are shown in \Fig{tov_freqs_flashx}, and we observe that the effective potential $\Phi_\text{E}$ and the GR results have a fundamental mode of ${\sim}1.2$ kHz, while \code{GREP} oscillates faster at ${\sim}2$ kHz.
These results demonstrate that both effective potentials are capable of reaching a new stably evolving state, but in both cases a state that differs from the full GR result: while our new potential matches the fundamental frequency of the oscillations, the amplitude of the oscillations is too small, while for \code{GREP}, the fundamental frequency is too high while the amplitude matches the full GR result well.

\section{Conclusion}\label{sec:conclusion}

We have introduced a new method for including a GR effective potential in Newtonian CCSN simulations.
A co-moving frame projection mirrors the TOV equations that describe the structure of a spherically symmetric relativistic star, while an Eulerian projection takes a form similar to the radial-gauge polar-slicing formulations used in some GR codes, e.g., \code{GR1D}.
To facilitate replacing the monopole term in the full Newtonian gravitational potential, we have further adapted the effective potential into a `Newtonian + GR correction' form.

We have implemented this new effective potential in both the \code{Chimera} and \code{Flash-X} codes, and have tested both on a series of collapse simulations to compare both Newtonian and other effective potentials, as well as with the full GR \code{GR1D} and \code{AGILE-BOLTZTRAN} simulations.
For adiabatic collapse simulations, we find close agreement of the new effective potential with the GR results for the post-bounce evolution, which are qualitatively different from the \code{GREP} results.
CCSN simulations show similar results for the post-bounce dynamics, e.g., faster expanding shocks compared to other effective potentials, but differences in numerical methods complicate any comparisons with the results from fully GR codes.
Evolving an isolated neutron star with different effective potentials shows similar dynamics as those seen in the full GR evolution in \code{SphericalNR}.

Approximating GR for CCSN simulations using \code{Chimera} and \code{Flash-X} proves a useful addition to these codes' capabilities.  Multi-dimensional \code{Chimera} simulations examining the impact of the new effective potential are currently underway and will provide a new baseline for future \code{Chimera} models. 
Extending the $\order{v/c}$ neutrino radiation transport in \code{thornado} \citep{Endeve2026} to include a dependence on the effective potential via the lapse and its derivatives will further augment combined \code{Flash-X+thornado} multidimensional CCSN simulations.

\begin{acknowledgements}
The authors would like to thank E. Endeve for many insightful conversations.
This work at Oak Ridge National Laboratory is supported under contract DE-AC05-00OR22725 with the U.S. Department of Energy.
This work was supported in part by the U.S. Department of Energy, Office of Science, Office of Advanced Scientific Computing Research and Office of Nuclear Physics, Scientific Discovery through Advanced Computing (SciDAC) program.
This research used resources of the Oak Ridge Leadership Computing Facility at the Oak Ridge National Laboratory, which is supported by the Advanced Scientific Computing Research programs in the Office of Science of the U.S. Department of Energy under Contract No. DE-AC05-00OR22725.
This research used resources of the National Energy Research Scientific Computing Center (NERSC), a Department of Energy User Facility using NERSC award NP-ERCAP0032339.
This research used resources from the Texas Advanced Computing Center's (TACC) Frontera and Vista supercomputer allocations (award PHY20010).
\end{acknowledgements}

\software{\code{Chimera} \citep{Bruenn2020}, \code{Flash-X} \citep{Dubey2022}, \code{GR1D} \citep{OConnor2010,OConnor2015}, \code{AGILE-BOLTZTRAN} \citep{Liebendorfer2004}, \code{SphericalNR} \citep{Mewes2018,Mewes2020}, \code{matplotlib} \citep{Hunter2007}, \code{numpy} \citep{Harris2020}, \code{sympy} \citep{Meurer2017}}

\bibliography{EffectivePotentials}
\bibliographystyle{aasjournalv7}

\appendix

\section{Effective potential derivation}\label{app:derivations}

The Ricci tensor in \Eq{einstein_eq} results from the contraction
\begin{align}\label{eq:Rab}
    \tensor{R}{_{ab}} = \tensor{R}{^c_{acb}},
\end{align}
where the Riemann tensor is given by \citep{misner2017}
\begin{align}\label{eq:Rabcd}
    \tensor{R}{^a_{bcd}} = \pd{_c}\CC{a}{b}{d} - \pd{_d}\CC{a}{b}{c} + \CC{a}{e}{c}\CC{e}{b}{d} - \CC{a}{e}{d}\CC{e}{b}{c}.
\end{align}
The Christoffel symbols in \Eq{Rabcd} can be found from
\begin{align}\label{eq:christoffel}
    \CC{a}{b}{c} = \frac{1}{2}\tensor{g}{^{ad}}\qty(\pd{_b}\tensor{g}{_{cd}} + \pd{_c}\tensor{g}{_{bd}} - \pd{_d}\tensor{g}{_{bc}}),
\end{align}
of which the non-zero symbols for the metric describing the line element in \Eq{tov_ds2} are
\begin{equation}\label{eq:tov_christoffel}
    \begin{alignedat}{4}
    \CC{t}{t}{r} &= \pd{_r}\ln\alpha \qc &\CC{r}{t}{t} &= \alpha^2 X^{-2} \pd{_r}\ln\alpha,\\
    \CC{r}{r}{r} &= \pd{_r}\ln X \qc &\CC{r}{\theta}{\theta} &= -r X^{-2},\\
    \CC{r}{\varphi}{\varphi} &= -r X^{-2}\sin^2\theta \qc &\CC{\theta}{r}{\theta} &= \CC{\varphi}{r}{\varphi} = r^{-1},\\
    \CC{\theta}{\varphi}{\varphi} &= -\sin\theta\cos\theta\qc &\CC{\varphi}{\theta}{\varphi} &= \cot\theta.
    \end{alignedat}
\end{equation}
With these, the non-zero Ricci tensor components are
\begin{equation}\label{eq:tov_ricci_tensor}
\begin{aligned}
    \tensor{R}{_{tt}} &= \alpha^2 X^{-2}\qty[\pd{_r}\pd{_r}\ln\alpha + \qty(\pd{_r}\ln\alpha)\qty(\pd{_r}\ln\alpha) - \qty(\pd{_r}\ln\alpha)\qty(\pd{_r}\ln X) + 2r^{-1}\pd{_r}\ln\alpha],\\
    \tensor{R}{_{rr}} &= \qty(\pd{_r}\ln\alpha)\qty(\pd{_r}\ln X) - \pd{_r}\pd{_r}\ln\alpha - \qty(\pd{_r}\ln\alpha)\qty(\pd{_r}\ln\alpha) + 2r^{-1}\pd{_r}\ln X,\\
    \tensor{R}{_{\theta\theta}} &= \sin^{-2}\theta \tensor{R}{_{\varphi\varphi}} = X^{-2}\qty[r\pd{_r}\ln X - r\pd{_r}\ln\alpha + X^2 - 1].
\end{aligned}
\end{equation}
The Ricci scalar is then
\begin{align}\label{eq:tov_ricci_scalar}
    R &= 2X^{-2}\qty[\qty(\pd{_r}\ln\alpha)\qty(\pd{_r}\ln X) - \pd{_r}\pd{_r}\ln\alpha - \qty(\pd{_r}\ln\alpha)\qty(\pd{_r}\ln\alpha)]\nonumber\\
    &\quad + 2r^{-2}X^{-2}\qty[2r\qty(\pd{_r}\ln X - \pd{_r}\ln\alpha) + \qty(X^2 - 1)].
\end{align}

Assuming a stationary solution, the four-velocity is simply
\begin{align}\label{eq:tov_u}
    \tensor{u}{_a} = \qty(-\alpha,0,0,0) \qc \tensor{u}{^a} = \qty(\alpha^{-1},0,0,0),
\end{align}
and the projection operator orthogonal to $\tensor{u}{^a}$ is
\begin{align}\label{eq:tov_h}
    \tensor{h}{_{ab}} = \tensor{g}{_{ab}} + \tensor{u}{_a}\tensor{u}{_b}.
\end{align}
The projection of the left-hand side of \Eq{einstein_eq} along the four-velocity is
\begin{align}\label{eq:uuG}
    \tensor{u}{^a}\tensor{u}{^b}\qty(\tensor{R}{_{ab}} - \frac{1}{2}R\tensor{g}{_{ab}}) &= \alpha^{-2}\tensor{R}{_{tt}} + \frac{1}{2}R = r^{-2}X^{-2}\qty[2r\pd{_r}\ln X + \qty(X^2 - 1)],
\end{align}
while the projection of the stress-energy tensor in \Eq{Tab} defined in terms of and along the four-velocity in \Eq{tov_u} is simply
\begin{align}\label{eq:uuT}
    \tensor{u}{^a}\tensor{u}{^b}\tensor{T}{_{ab}} = \rho + \mathcal{J}.
\end{align}
Using \Eqns{uuG}{uuT} in \Eq{einstein_eq} gives the result in \Eq{dX_dr}.
Similarly, the $rr$-component of the orthogonal projection of the left-hand side of \Eq{einstein_eq} gives
\begin{align}\label{eq:hhG}
    \tensor{h}{^a_r}\tensor{h}{^b_r}\qty(\tensor{R}{_{ab}} - \frac{1}{2}R\tensor{g}{_{ab}}) &= \tensor{R}{_{rr}} - \frac{1}{2}R\tensor{g}{_{rr}} = 2r^{-2}\qty[r\pd{_r}\ln\alpha - \qty(X^2 - 1)],
\end{align}
and the $rr$-component of the orthogonal projection of the stress-energy tensor in \Eq{Tab} defined with the four-velocity in \Eq{tov_u} is
\begin{align}\label{eq:hhT}
    \tensor{h}{^a_r}\tensor{h}{^b_r}\tensor{T}{_{ab}} = X^2\qty(p + \mathcal{K}).
\end{align}
Using \Eqns{hhG}{hhT} in \Eq{einstein_eq} gives the result in \Eq{dalpha_dr}.

In the frame of an Eulerian observer with a four-velocity
\begin{align}\label{eq:n_E}
    \tensor{n}{_a} = \qty(-\alpha,0,0,0) \qc \tensor{n}{^a} = \qty(\alpha^{-1},0,0,0),
\end{align}
we decompose the co-moving observer's four-velocity into components tangential to $\tensor{n}{^a}$ and to the fluid velocity $\tensor{v}{^a}$
\begin{align}\label{eq:u_E}
    \tensor{u}{^a} = W\qty(\tensor{n}{^a} + \tensor{v}{^a}),
\end{align}
where $W = -\tensor{n}{_a}\tensor{u}{^a} = 1/\sqrt{1-v^2}$ is the Lorentz factor and $\tensor{n}{_a}\tensor{v}{^a} = 0$.
The projection operator orthogonal to $\tensor{n}{^a}$, which also acts as the spatial metric in the frame of the Eulerian observer, is then
\begin{align}\label{eq:gamma_E}
    \tensor{\gamma}{_{ab}} = \tensor{g}{_{ab}} + \tensor{n}{_a}\tensor{n}{_b}.
\end{align}
The projection of the left-hand side of \Eq{einstein_eq} along the Eulerian four-velocity is
\begin{align}\label{eq:nnG}
    \tensor{n}{^a}\tensor{n}{^b}\qty(\tensor{R}{_{ab}} - \frac{1}{2}R\tensor{g}{_{ab}}) &= \alpha^{-2}\tensor{R}{_{tt}} + \frac{1}{2}R = r^{-2}X^{-2}\qty[2r\pd{_r}\ln X + \qty(X^2 - 1)],
\end{align}
while the projection of the stress-energy tensor in \Eq{Tab} defined in terms of and along the Eulerian four-velocity in \Eq{u_E} is
\begin{align}\label{eq:nnT}
    \tensor{n}{^a}\tensor{n}{^b}\tensor{T}{_{ab}} = \rho_0hW^2 - p + \mathcal{E}.
\end{align}
Using \Eqns{nnG}{nnT} in \Eq{einstein_eq} gives the result in \Eq{dX_dr_E}.
Similarly, the $rr$-component of the orthogonal Eulerian projection of the left-hand side of \Eq{einstein_eq} gives
\begin{align}\label{eq:ggG}
    \tensor{\gamma}{^a_r}\tensor{\gamma}{^b_r}\qty(\tensor{R}{_{ab}} - \frac{1}{2}R\tensor{g}{_{ab}}) &= \tensor{R}{_{rr}} - \frac{1}{2}R\tensor{g}{_{rr}} = 2r^{-2}\qty[r\pd{_r}\ln\alpha - \qty(X^2 - 1)],
\end{align}
and the $rr$-component of the orthogonal Eulerian projection of the stress-energy tensor in \Eq{Tab} defined with the four-velocity \Eq{u_E} is
\begin{align}\label{eq:ggT}
    \tensor{\gamma}{^a_r}\tensor{\gamma}{^b_r}\tensor{T}{_{ab}} = X^2\qty(\rho_0hW^2v^2 + p + \mathcal{P}).
\end{align}
Using \Eqns{ggG}{ggT} in \Eq{einstein_eq} gives the result in \Eq{dalpha_dr_E}.

\restartappendixnumbering
\section{\code{GREP} Implementation}\label{app:grep}

The `Case A' effective potential of \citet{Marek2006} (\code{GREP}) as implemented in \code{Chimera} and \code{FLASH}/\code{Flash-X} takes the form
\begin{gather}
    \Phi_\text{GREP}(r) = \int_\infty^r\dd{r'}\qty[\frac{m_\text{GREP}}{r'^2} + 4\pi r'\qty(p + \mathcal{K})]\frac{h}{\Gamma^2}\label{eq:phi_A},\\
    m_\text{GREP}(r) = 4\pi\int_0^r\dd{r'}r'^2\qty(\rho + \mathcal{J} + \frac{v\mathcal{H}}{\Gamma})\Gamma\label{eq:m_A},\\
    \Gamma = \sqrt{1 + v^2 - \frac{2m_\text{GREP}}{r}}\label{eq:Gamma_A},
\end{gather}
where $v$ is the radial velocity and $\mathcal{H}$ is the neutrino radial flux.
The primary differences between this effective potential and the one presented in \sref{formulation} are the presence of velocity-dependent terms in the mass $m_\text{GREP}$ and metric function $\Gamma$, and the additional factor of the specific enthalpy $h$ in the integrand of the potential.
The extra factor of $\Gamma$ in the mass integrand results from the `Case A' modification of the TOV-like mass equation.
Replacing the Newtonian monopole term then gives the effective potential as
\begin{align}\label{eq:Phi_grep}
    \bar{\Phi}_\text{GREP}(\vb{r}) = \Phi_N(\vb{r}) - \Phi_0(r) + \Phi_\text{GREP}(r).
\end{align}

\end{document}